\documentclass[conference]{IEEEtran}
\IEEEoverridecommandlockouts
\usepackage{mathtools}
\usepackage[utf8]{inputenc}
\usepackage{fancyhdr} 
\usepackage{balance}
\usepackage{cite}
\usepackage{amsmath,amssymb,amsfonts}
\usepackage{algorithmic}
\usepackage{graphicx}
\graphicspath{{image/}}
\usepackage{textcomp}
\usepackage{xcolor}
\usepackage[english]{babel}
\usepackage{multirow}
\usepackage{multicol}
\usepackage{url}
\usepackage[autostyle]{csquotes}

\usepackage{tikz}
\usetikzlibrary{positioning}

\def\BibTeX{{\rm B\kern-.05em{\sc i\kern-.025em b}\kern-.08em
    T\kern-.1667em\lower.7ex\hbox{E}\kern-.125emX}}
\begin{document}

\title{Multicriteria Recommendation System by Leveraging Predefined, Implicit, and Undefined Criteria}

\author{
Emrul Hasan\\
 Department of Computer Science\\
 Toronto Metropolitan University\\
 Toronto, Canada\\
  \texttt{e1hasan@torontomu.ca}
  \and
Chen Ding\\
 Department of Computer Science\\
 Toronto Metropolitan University\\
 Toronto, Canada\\
  \texttt{cding@torontomu.ca}
}

\maketitle

\begingroup\renewcommand\thefootnote{}\footnotetext{This preprint has not undergone peer review or any post-submission improvements or corrections. The Version of Record of this contribution is published in Social Networks Analysis and Mining (ASONAM 2024), Lecture Notes in Computer Science, Springer, and is available online at https://doi.org/10.1007/978-3-031-78538-2\_6.}\addtocounter{footnote}{-1}\endgroup

\textbf{Abstract}--
Data sparsity, a long-standing problem in the recommendation system, occurs when the user-item utility matrix is sparse. For years, many approaches have been experimented with, including utilizing user's textual reviews and multiple criteria ratings. In a review-based recommendation system, user preferences and item attributes are learned from reviews, while in multi-criteria recommendation systems, criteria ratings are used for the recommendation task. Multi-criteria are domain-specific pre-defined explicit criteria and users are bound to provide feedback only on these criteria with discrete values. On the contrary, users are free to express their opinions through textual reviews, which can cover multiple perspectives or aspects. It is possible that these pre-defined, domain-specific explicit criteria are not enough to fully capture the multi-faceted user preferences. To this end, we propose a novel multi-criteria recommendation model that utilizes predefined, implicit, and undefined criteria. We use a semantic similarity-based sentence clustering method to identify the predefined and implicit criteria and a sentiment analyzer to estimate their ratings. Semantic similarity between each sentence in the review and the predefined criteria are calculated, and then the sentence is assigned to the most similar criteria. The criteria that are extracted from the review and are aligned with predefined criteria are referred to as implicit criteria. A sentence is considered as expressing opinions on an undefined criterion if the similarity score between this sentence and all the predefined criteria is lower than a predefined threshold. Ratings are computed for each extracted implicit criterion and the undefined criterion based on the review content. Finally, we use all three types of criteria and an aggregation model to make the final rating prediction for the recommendation system.  Our proposed method demonstrates the superiority compared to several baselines on TripAdvisor and Beer Advocate datasets.


\section{Introduction}
Due to the rapid growth of online data, recommendation system has become a ubiquitous tool for e-commerce and content providers such as Amazon, Netflix, YouTube, etc. These systems address the challenge of information overload by filtering content according to user preferences. At its core, these systems aim to learn user preferences and item attributes by leveraging historical interaction data. Traditional recommendation systems mainly utilize user-item rating data. Collaborative Filtering (CF) \cite{sarwar2001item} is one type of the most successful and widely used recommendation models. A popular CF model is matrix factorization (MF) \cite{koren2009matrix}, which learns the underlying common factors behind the user ratings.

While traditional collaborative filtering methods such as MF are effective, they fall short in modeling complex user-item interactions, particularly in scenarios where the rating matrix is highly sparse \cite{natarajan2020resolving}. Additionally, since CF utilizes single-criterion (overall rating) to produce recommendation results, they cannot capture fine-grained user preferences, which could lead to low performance.

\begin{figure}[ht]
\centering
\includegraphics[width=0.5\textwidth]{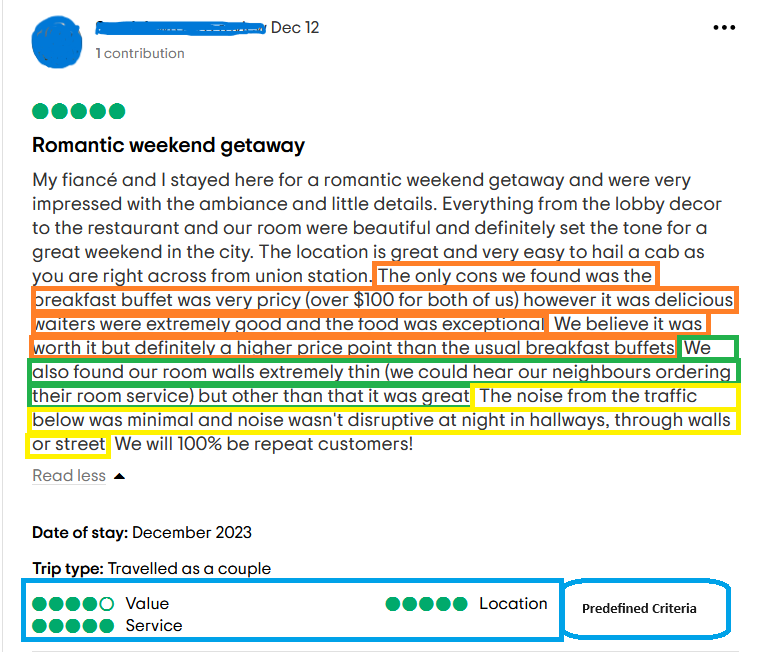}
    \caption{A hotel review collected from tripadvisor.com representing review, criteria ratings, and the overall ratings}
\label{fig:criteria}
\end{figure}

When users choose a product or service, they typically consider one or multiple attributes rather than relying solely on overall ratings. In the context of hotels, a user might prioritize factors such as location and service, or focus solely on price. The overall rating could be high due to previous users' satisfaction on other criteria that this user may not find relevant (see Figure \ref{fig:criteria}), leading to a situation where the user may not receive appropriate recommendations. To address this issue, the researchers turned their attention to multi-criteria recommendation systems where multiple criteria ratings are utilized to learn user preferences \cite{nassar2020multi, shambour2021deep,hong2021multi}. Another direction is to build aspect-based recommendation systems where user preferences and item attributes are learned based on different aspects that users discussed in the review text\cite{chin2018anr,li2021aspect,wu2019arp}. 

While aspect-based and multi-criteria rating-based recommendations rely on a single data source (review or criteria) for user preference learning, we argue that a more accurate representation of user preferences can be learned when we consider both the criteria ratings and the sentiment expressed through reviews. It is observed that in addition to pre-defined criteria (domain-specific criteria that are defined by the business), users express their opinions on different aspects more freely through reviews. For example, in Figure \ref{fig:criteria}, predefined criteria are value, location, and service, however, the review reveals additional aspects such as breakfast and noise. Therefore, it is important to consider both the review and the criteria ratings while learning user preferences and item characteristics. 

To this end, we propose a novel multi-criteria recommendation model that utilizes both the predefined criteria and the criteria (aspects) extracted from the review. The predefined criteria may also be discussed in the review, however, the ratings (sentiment scores) associated with them may not be the same as the ratings users explicitly give to the predefined criteria. So, to differentiate the two, we call the ratings based on the review implicit criteria ratings. If the aspects extracted from the review is not available in the predefined criteria rating list, they are named undefined criteria.  

Inspired by \cite{nassar2020novel}, we leverage an aggregation-based multi-criteria recommendation model. The aggregation-based  \cite{nassar2020multi} method aims to learn a function that captures the relationship between criteria ratings and the overall ratings. Our proposed method comprises four modules: 1) criteria extraction, 2) criteria rating prediction, 3) aggregation function learning, 4) overall rating prediction.

We assume that each sentence or sub-sentence in the review contains user's opinion on one criterion. Some criteria are discussed in multiple sentences. To extract the criteria from the review, we split the review into sentences and cluster them by comparing the semantic similarity between predefined criteria and the split sentences. The sentences that have high similarity degrees with predefined criteria are clustered into different homogeneous groups, each of which corresponds to a predefined criterion. The implicit rating scores are calculated using sentiment analyzer \cite{yao2019automated}. The sentences that have a low similarity score with any predefined criteria are recorded in a separate cluster and named as undefined criteria. Both criteria rating prediction and aggregation function modules are adopted from the paper introduced by Nassar et al. \cite{nassar2020multi}  where they use the Deep Neural Network and matrix factorization for criteria rating prediction and deep neural network for aggregation function.

The following are the major contributions of our research:

\begin{enumerate}

    \item We propose a novel multi-criteria recommendation model by leveraging both reviews and the predefined criteria. We show that the user's overall preference on an item not only depends on the predefined criteria ratings but also on the ratings extracted from reviews.

   \item  We leverage all three types of criteria ratings to learn the overall rating - predefined (explicit ratings given on predefined criteria), implicit (rating extracted from reviews on aspects that can match with predefined criteria), and undefined (rating extracted from reviews on aspects that do not match with predefined criteria). We extract implicit and undefined criteria from the review by using a semantic similarity-based clustering method. To the best of our knowledge, our cluster-based aspect extraction method is new, and the process of aligning (similarity-based matching) the criteria extracted from reviews with the predefined criteria to get undefined criteria ratings and include them in the overall rating prediction is also considered novel.

   \item We assess the performance of the model with publicly available Tripadvisor and Beeradvocate datasets and compare the results with several state-of-the-art.

\end{enumerate}

In Section 2, a comprehensive literature review is presented. Section 3 offers a thorough description of the problem and the model. Empirical studies and their results are detailed in Sections 4 and 5, respectively, while Section 6 encompasses the conclusion and outlines future work.

\section{Related Work}

The multi-criteria recommendation system leverages multiple criteria ratings provided by users to predict overall ratings, demonstrating enhanced accuracy compared to the single-criteria rating-based approach.  These techniques are broadly categorized into two groups: memory-based methods and model-based methods. Similarity-based methods, also called memory-based methods try to compute the similarity independently for each criteria rating using traditional methods, followed by aggregating these values into a single similarity using techniques like averaging \cite{adomavicius2007new}, and weighted sums of individual similarities \cite{tang2009pedagogical}. The alternative method involves employing multi-dimensional distance measures, such as Euclidean and Manhattan distance, for computing the distances between different criteria ratings. The model-based method uses different recommendation algorithms including MSVD\cite{de2000multilinear}, SVR\cite{awad2015support}, and aggregation-function methods \cite{grabisch2009aggregation}. 

Nassar et al. \cite{nassar2020multi} introduce an aggregation-function-based recommendation by leveraging a deep neural network (DNN) and matrix factorization (MF). The model consists of two parts, criteria ratings prediction using a deep collaborative filtering model by leveraging DNN and MF, and learning an aggregation function using a DNN. Finally, the unknown rating is predicted using the learned aggregation function and the predicted multi-criteria ratings. Hong et al.\cite{hong2021multi} develop a multi-criteria tensor model for tourism recommender systems. They explore user preferences by incorporating cultural factors into the evaluation of multiple criteria ratings. The paper introduces two models referred to as "single tensor" models, which encompass users (or countries), items, multi-criteria ratings, and cultural groups. The objective is to simultaneously address the inherent structure and interrelationships among these factors within the recommendation process. 

It is demonstrated that recommendation performance is influenced by multiple factors. Wang et al.\cite{Wangetal2013} introduce a recommendation system with a Hybrid Deep Tensor Decomposition model, incorporating multiple criteria. It is observed that the consideration of multiple factors has an impact on recommendation performance. In this case, side information is processed with deep structure, and 3D user-item criteria ratings are processed with tucker decomposition \cite{kim2007nonnegative}. Hong et al. \cite{hong2021clustptf} introduced ClustPTF (Cluster-based Parallel Tensor Factorization), which employs sentiment analysis to mitigate data sparsity within the tensor model \cite{kim2007nonnegative}. Next, K-means clustering is used to cluster similar user preferences, thereby improving recommendation diversity. Shambour \cite{shambour2021deep} introduces a deep learning-based multi-criteria recommendation system. The system employs a deep auto-encoder to unveil latent user preferences for an item. Although all the above-mentioned multi-criteria recommendation models have proven to be successful in performance improvement, they use predefined criteria for overall preference learning, ignoring the user sentiment expressed through review.

On the other hand, aspect-based approaches concentrate on learning the aspect representations from reviews and utilizing them to predict overall preferences. ANR \cite{chin2018anr} introduces an Aspect-based Neural Recommendation model incorporating an attention mechanism to learn user and item representations simultaneously. Similarly, Li et al. \cite{li2021aspect} develop an aspect-based fashion recommendation system that extracts both local and global features to estimate overall ratings. Since relying on explicit criteria ratings, aspect-based recommendation systems face the challenge that inferred ratings from reviews may not always be accurate. Again, aspect-based recommendation considers only review for preference learning, ignoring the explicit multiple criteria ratings. 

Because of the deficiency of the review-only and multi-criteria only recommender systems, in cases where both reviews and criteria ratings are accessible, such as on platforms like Tripadvisor.com, and BeerAdvocate.com, which gathers user feedback through textual reviews, criteria ratings, and overall ratings, it is beneficial to leverage all the information for preference learning. In a more recent development, Hasan et al. \cite{emrul2022mrrrec} propose a multi-criteria recommendation model that learns aspect representation from the review and then combines the aspect ratings with the explicit criteria ratings for the overall rating prediction.  

The most similar research to our proposed model are the ones by Nassar et al.\cite{nassar2020multi} and by Hasan et al. \cite{emrul2022mrrrec}. The major difference between our work and the former is that they solely rely on criteria ratings without considering textual reviews. In contrast, our model integrates textual reviews and criteria ratings for a more comprehensive analysis. The difference between our work and the latter is that they did not consider the alignment of the extracted aspects with predefined criteria.

\section{Proposed Model}

The proposed method comprises four modules: 1) criteria extraction, 2) criteria rating prediction, 3) aggregation function learning, and 4) overall rating prediction. Each of the components is briefly discussed in the following subsections.
\subsection{Preliminaries}
\begin{figure}
\centering
\includegraphics[width=0.5\textwidth]{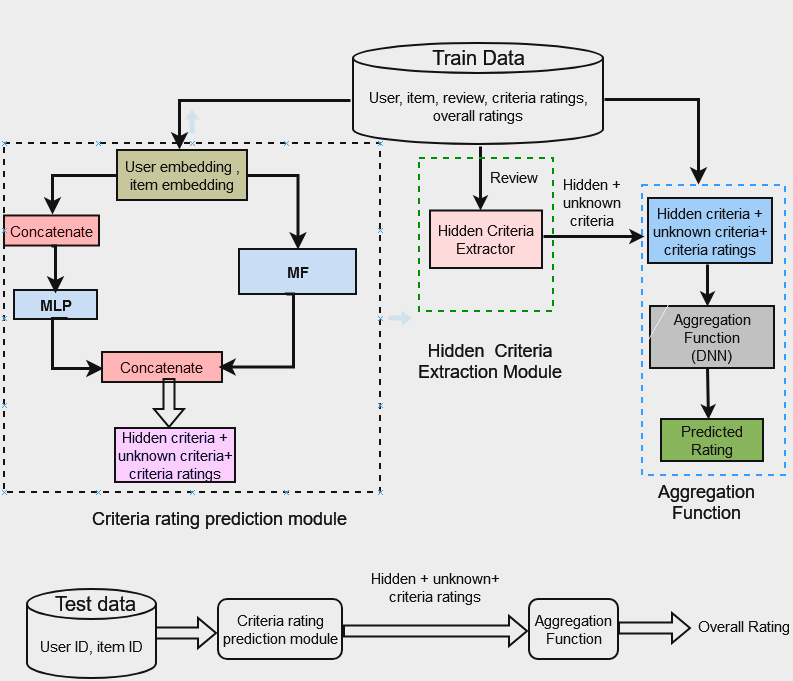}
    \caption{The detailed architecture of the proposed model.}
\label{fig: architecture}
\end{figure}

Consider a user $u$ $\epsilon$ $U$ provided feedback to an item $i$ $\epsilon$ $I$ with three different modes: review $x$ $\epsilon$ $X$, criteria rating $r_{K}$ $\epsilon$ $r_K^N$, and overall rating $r_{0}$ $\epsilon$ $R$  where $K$ and $N$  are the number of predefined criteria, and the total number of reviews respectively. Our first aim is to extract the implicit and undefined criteria and corresponding ratings from the review. In the second step, our goal is to predict the criteria ratings using Neural collaborative filtering. In the third step, we aim to learn an aggregation function using all criteria (undefined, implicit, and predefined) and the overall rating. Finally, the overall rating for the unknown item is predicted using predicted criteria and the learned aggregation function. The detailed architecture of the model is shown in Figure \ref{fig: architecture}.

\subsection{Criteria extraction from the review}
\begin{figure}
\centering
\includegraphics[width=0.48\textwidth]{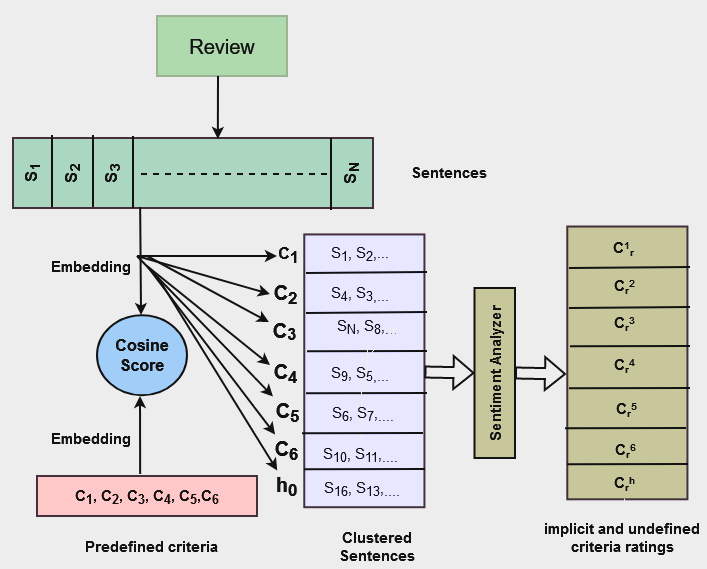}
    \caption{The detailed architecture of the proposed criteria extraction model.}
\label{fig:clustering}
\end{figure}

In this step of the process, we split the review into sentences on punctuation symbol \enquote{.} , and words such as \enquote{but}, \enquote{while}, \enquote{however}, \enquote{nonetheless}. Here we assume that when using these conjunctions users start to speak on different aspects. Next, we compute the semantic similarity between each sentence and the predefined criteria. We utilize the pre-trained sentence transformer (e.g. SBERT) to generate embeddings for both sentences and the predefined criteria. Then, cosine similarity between each criterion and the split sentences are computed.  Sentences are assigned to one of the predefined criteria if the similarity score with that criterion is bigger than a threshold value. Sentences whose similarity scores with all the predefined criteria are less than this threshold value are assigned to one cluster, corresponding to the undefined criteria. We consider these sentences express users' opinions on aspects that are not covered by predefined criteria. Once the clustering is done, all the sentences in the cluster for each criterion are concatenated, and finally, a sentiment analyzer is used to estimate the rating for each of the clusters. The detailed sentence clustering process is shown in Figure \ref{fig:clustering}.

The similarity score is measured using cosine similarity and mathematically the formula for vectors $P$ and $Q$ can be written as

\begin{eqnarray}
\text{Cosine Similarity}(P, Q) =\frac{P \cdot Q}{\|P\| \cdot \|Q\|}
\end{eqnarray}
$\|P\|$ and $\|Q\|$ represents the magnitude of $P$ and $Q$. The resulting value spans from -1 to 1, with -1 indicating complete dissimilarity and 1 indicating complete similarity

\subsection{Criteria Rating Prediction}
Inspired by Nassar et al. \cite{nassar2020multi}, we employ Neural Multi-criteria Collaborative Filtering for criteria rating prediction. The model consists of 3 steps, Matrix Factorization (MF), DNN, and fusion of MF and DNN. Since the IDs are categorical features, motivated by He et al. \cite{he2017neural}, the user ID and item ID are vectorized by leveraging the embedding technique and initialized with random values. Consider the embedding for user $u$ and item $i$ are $e_{u}$ and $e_{i}$ respectively. The MF model takes the element-wise product as input, whereas the DNN model takes a concatenated vector as input and these can be expressed as follows. 
For MF, input vector is
\begin{eqnarray}
E = f(e_{u},e_{i})=e_{u} \odot e_{i}
\end{eqnarray}
and the output is represented as 
\begin{eqnarray}
\hat y = \sigma(w^T(e_{u} \odot e_{i}))
\end{eqnarray}

where $\sigma$ is the activation function and $w$ is the weight. For DNN, input vector is the concatenation of $e_{u}$ and $e_{i}$
\begin{eqnarray}
E=f_{1}(e_{u},e_{i}) =\begin{bmatrix}
    e_{u}\\
     e_{i}
\end{bmatrix}
\end{eqnarray}

The activation function employed is Rectified Linear Unit (ReLU)\cite{nair2010rectified}, chosen for its recognized efficiency in various applications
\begin{eqnarray}
\sigma=ReLU (z)=max(0,z)
\end{eqnarray}

The hidden layers for the MLP can be written as

\begin{eqnarray}
f_{2}(f_{1}) =\sigma(W_{2}^T f_{1} +b_{2})
\end{eqnarray}

where $W_{2}$, $b_{2}$ and $f_{1}$ represent weights, bias and hidden layer 1 respectively.

For $L$ number of layers the formula can be generalized as 
\begin{eqnarray}
f(E_{L-1}) =\sigma_{L}(W_{L}^T E_{L-1} +b_{L})
\end{eqnarray}

The final output of the DNN can be written as
\begin{eqnarray}
\hat y =\sigma(h^T(f_{L}(E_{L}-1))
\end{eqnarray}

where L is the number of layers. $h$ is the weight matrix.

Now, we need to defuse MF and MLP so that the model can learn complex user-item interaction. The step of combining the output from two modules can be written as 

\begin{eqnarray}
\hat y_{ui} =\sigma\left( W^{T}(e_{u} \odot e_{i}) + h^T\begin{bmatrix}
    e_{u}\\
     e_{i}
\end{bmatrix} +b\right)
\end{eqnarray}

To enhance the flexibility of the fused model, we enable MF and DNN to learn distinct embeddings. The combination of these two models is achieved by concatenating their last hidden layers. Figure \ref{fig: architecture} depicts our proposed approach, and its formulation can be simplified as follows

\begin{eqnarray}
\hat y_{ui} =\sigma\left(h^T \begin{bmatrix} f^{MF}\\ f^{MLP}\end{bmatrix} \right)
\end{eqnarray}

where $f^{MF}$ and $f^{MLP}$ are the final layer of MF and MLP respectively. $\sigma$ is the activation function, e.g. $ReLU$

\begin{eqnarray}
f^{MF} = e_{u}^{MF} \odot e_{i}^{MF}
\end{eqnarray}
where $e_{u}^{MF}$ and $e_{i}^{MF}$ are the user and item embeddings respectively for MF model.
Similarly,
\begin{eqnarray}
f^{MLP} = \sigma_{L}\left(W_{L}^T(...\sigma_{2}(W_{2}^T \begin{bmatrix}
    e_{u}^{MLP}\\
     e_{i}^{MLP}\end{bmatrix} +b_{2} )...\right)
\end{eqnarray}
where $e_{u}^{MLP}$ and $e_{i}^{MLP}$ are the user and item embedding.

\subsection{Learning Aggregation Function}
Since the overall rating is the decomposition of predefined, undefined, and implicit criteria ratings, the relationship between them and the overall ratings can be learned using an aggregation function.  We employ a deep neural network as the aggregation function to capture the relationship between the criteria ratings and the overall ratings. Mathematically, this relationship can be expressed as follows:
\begin{eqnarray}
r_{o}=f(r_{1}, r_{2},...,r_{K})
\end{eqnarray}
where $k$ is the number of criteria, and $r_{0}$ is the overall rating.

The network takes the criteria ratings as input and produces the overall ratings as output. Considering the sensitivity of DNN to input distribution and scaling \cite{ioffe2015batch}, we normalize the features. The normalization of a sample $r_{i}$ can be done as follows:

\begin{eqnarray}
Z_{i}=\frac{r_{i} -m}{s}
\end{eqnarray}

where $s$ represents the standard deviation of the training samples. Therefore the input can be formulated as follows: 
\begin{eqnarray}
\chi=[z_{1}, z_{2},....,z_{K}]^T
\end{eqnarray}
where $K$ is the number of criteria.
The final input vector is then passed through several dense layers followed by a $ReLU$ layer and output can be represented as 
\begin{eqnarray}
r_{0}=y_{ui}
\end{eqnarray}
$r_{0}$ denotes the overall rating for user $u$ to an unknown item $i$.
\subsection{Overall Rating prediction}

After training the model and learning the function, the next step is to predict the overall rating for test users. This involves predicting criteria ratings and then feeding them into the aggregation function to determine the overall ratings.

The predicted ratings are used as input to the learned function $f(r)$ to predict the final output and are defined as

\begin{eqnarray}
\hat r_{o}=f( \hat r_{1}, \hat r_{2},..., \hat r_{K} )
\end{eqnarray}

where $ \hat r_{o}$ is the final overall rating and $\hat r_{1},\hat r_{2},...,\hat r_{K}$ the predicted criteria ratings. 
\subsection{Objective Function}
To learn the parameters of both DNN models (overall rating and the criteria rating), we exploit the regression with squared loss as the objective function

\begin{eqnarray}
J=\sum (\hat r_{0}-r_{0})^2 + \lambda_{\theta} \|\Theta\|^2
\end{eqnarray}
$r_{0}$ is the real rating of
user $u$ for item $i$,
$\hat r\_{0}$ is the predicted rating, $\Theta$ denotes all the parameters. 
$\lambda_{\theta}$$\|\Theta\|^2$ is used as regularization to prevent the model from overfitting. Our goal is to minimize the objective function $J$. The model is trained through a backpropagation technique. For a fair comparison, we utilize Adam \cite{kingma2014adam} optimizer and MSE as the loss function.

\section{Experiments}
\subsection{Datasets}
We use two publicly available multicriteria datasets: Tripadvisor\footnote{\url{https://www.cs.virginia.edu/~hw5x/Data/LARA/TripAdvisor/}} and BeerAdvocate\footnote{\url{https://cseweb.ucsd.edu/~jmcauley/datasets.html\#multi\_aspect}}. The reason we choose these datasets is that they come with three forms of feedback, e.g. textual review, criteria ratings, and overall rating. Tripadvisor.com is a famous hotel and tourism management platform where users can book hotels, flights, and tourist attractions. The TripAdvisor dataset was gathered in 8 years, from 2004 to 2012. The scores for both individual criteria and overall ratings span a scale from -1 to 5. Ratings below 1 and criteria with over 75\% missing values are eliminated. In terms of data cleaning, for our model, we remove only unwanted characters and URLs, whereas for a bag of word embedding techniques for vectorization, we clean the data extensively, e.g. we eliminate the stopwords, punctuations, and newline characters. All reviews are tokenized using the Natural Language Processing Toolkit (NLTK) [55], and a vocabulary is constructed comprising the top 50,000 most frequent words. When using sentence BERT for vectorization, we do not eliminate punctuations and stopwords because of their contextual meaning. BeerAdvocate is a beer review platform that provides textual reviews, and ratings on individual criteria such as $appearance$, $aroma$, $palate$, $taste$, and $overall$. The ratings for appearance and palate are available on a scale of 1 to 5 whereas for aroma and taste are on a scale of 1 to 10, and overall ratings are collected out of 20. We normalized all the ratings to a scale of 5.  Since we split the reviews into sentences and the number of sentences is large (e.g. for a sample subset with 30,000 reviews, the total number of sentences is 81530), the step to compare each sentence vector with the criterion vector could require high computation power.

Because of the review length difference, we follow two different strategies for the two datasets to obtain the subsets. After the general cleaning, we examine the review lengths and sentence length distribution plots. For the TripAdvisor dataset, it is seen that the minimum and maximum review lengths are 0 and 4702 respectively. From the number of sentence distribution plots, we can see that the majority of the reviews fall between 5 and 8 sentences, as shown in Figure 4. Therefore, we discard the reviews with the number of sentences smaller than 5 and larger than 10. Finally, we randomly sampled 30000 reviews from the cleaned dataset. For the BeerAdvocate dataset, the minimum and maximum review lengths are 0 and 972, respectively. Nevertheless, the lengths of majority of the reviews fall within the range of 10 to 200 tokens. Sentence splits reveal that  majority of the reviews have between 0 and 5 sentences, as shown in Figure \ref{fig:beerlen}. Therefore, we remove the reviews with the number of sentences greater than 5 and smaller than 4. The reason we choose 5 is that the number of predefined criteria for beer review is 4 and only a few reviews have sentences greater than 5. We randomly sampled a subset of 30000 reviews. 
\begin{figure}
\centering
\includegraphics[width=0.5\textwidth]{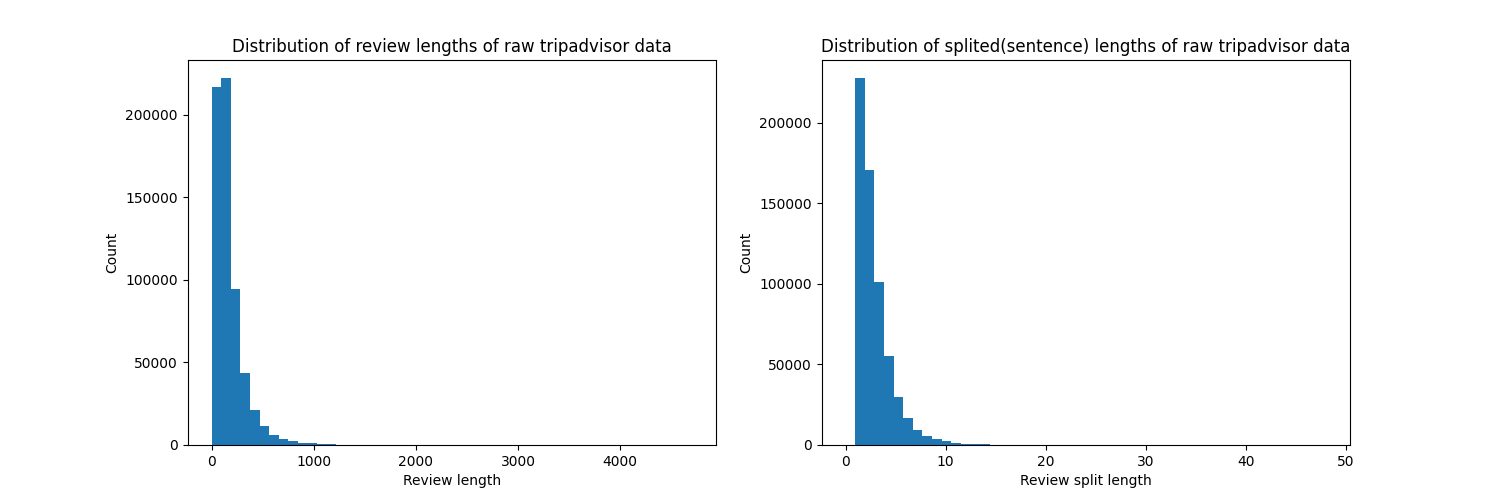}
    \caption{Review length and sentence length distribution for Tripadvisor dataset}
    \label{fig:triplen}
\end{figure}

\begin{figure}
\centering
\includegraphics[width=0.5\textwidth]{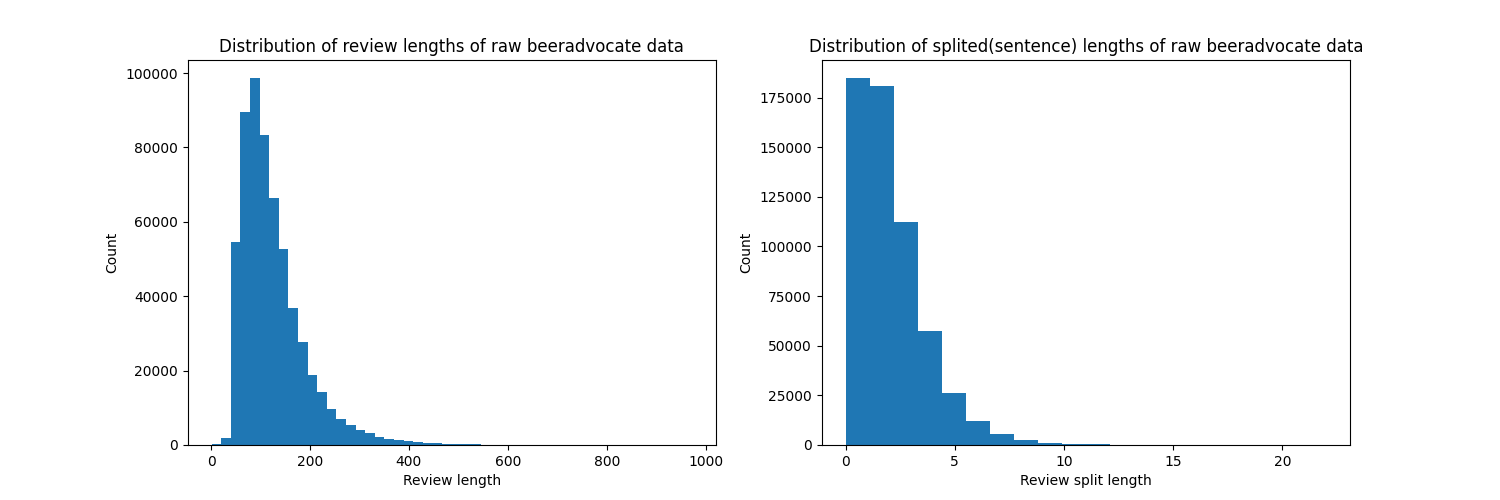}
    \caption{Review length and sentence length distribution for beer review dataset}

    \label{fig:beerlen}
\end{figure}

\subsection{Baselines}

We evaluate the effectiveness of our proposed model by comparing it with state-of-the-art review-based recommendation systems, including a multi-criteria recommendation system, an aspect-based recommendation system, and a general recommendation system. Our comparison is restricted to papers that have openly shared their code.

\begin{enumerate}
    \item \textbf{MRRRec} \cite{emrul2022mrrrec} is a multi-criteria recommendation model that leverages both the criteria ratings and the review to learn the user preferences. In this study, implicit criteria ratings are estimated from the textual review and then these ratings are concatenated with the explicit criteria (ratings that are predicted using a neural network from the user and item id embedding) ratings. Finally, the concatenated ratings are fed into a deep neural network to predict the overall rating.
    
    \item \textbf{ANR}\cite{chin2018anr}. In this study, user and item documents are constructed to learn the user preferences and item characteristics that govern the overall preferences on the item. An attention-based neural recommender system is introduced that detects the aspects and estimates their importance levels that users care for an item.
    
    \item \textbf{DeepCoNN} In \cite{he2017neural}, the Deep Cooperative Neural Network (DeepCoNN) is a state-of-the-art recommendation model that incorporates reviews for predicting ratings. The approach involves concatenating user and item representations, utilizing them as input to a Factorization Machine \cite{rendle2010factorization} to predict overall ratings.

    \item \textbf{Multi-criteria RS by Nassar et al. } \cite{nassar2020novel}. This represents a multi-criteria recommendation system based on deep learning. The model predicts overall criteria ratings through a two-step process. Initially, it predicts unknown $k$ multi-criteria ratings using a DNN, where user ID and item ID serve as features. In the subsequent step, the DNN model is employed once more to understand the relationship between multi-criteria ratings and overall ratings.

    \item  \textbf{NARRE} \cite{chen2018neural} employs attention mechanism to capture the representation of the user and item while predicting the overall ratings for the user.

    \item \textbf{DAML} \cite{liu2019daml} introduces a dual attention mutual learning method for recommendation by levering ratings and review. Review features are jointly learned by local and mutual attention, which can increase the interpretability.

    \item \textbf{MPCN} \cite{he2017neural} tries to improve the performance over the state-of-the-art D-ATTN \cite{seo2017interpretable} and DeepCoNN model \cite{zheng2017joint}. It considers that the selected reviews and items play an important role in preference learning. 
\end{enumerate}

We evaluate the performance of our model with two error metrics: MSE (Means Squared Error) and MAE (Mean Absolute Error). The reason behind choosing these two metrics is due to their widespread adoption in state-of-the-art rating prediction recommendation models\cite{liu2019daml, chin2018anr, deng2018neural, rafailidis2019adversarial}. We further extend our evaluation to precision, recall, and F1 accuracy measures. To compute the precision, recall, and F1, we formulate the problem as a binary classification problem by converting the ratings to 0 and 1 based on a threshold score. In our case, we set this threshold to 3.5. A rating above 3.5 is converted to 1; if it is less than 3.5, it is converted to 0.

\subsection{Experimental Settings}
 We partitioned the datasets into three subsets: training, validation, and test sets, using an 80, 10, 10 ratio. The DNN for both criteria rating and an aggregation function are trained and validated with training and validation datasets. We implemented our model and the baselines using Pytorch library and T4 GPU from Google Colab Pro. For a fair comparison, we kept the hyperparameters same as the baseline papers. In our experiments, for clustering tasks, we used different similarity threshold scores, ranging from 0.10 to 0.40 with an increment of 0.05. We leverage the Hugging Face \footnote{\url{https://huggingface.co/}} library to implement the Sentence Transformer.

\begin{table*}[t]

\centering
\caption{Performance comparison with baselines in terms of MSE and MAE}
\begin{center}
\begin{tabular}{p{3.0cm}p{3.1cm}p{3.1cm}}
\hline
\multicolumn{3}{c}{Performance Comparison} \\
\hline
Model & TripAdvisor & BeerAdvocate \\
\hline

& MSE {} {} MAE & MSE{}   {} MAE \\ 
 \hline
\textbf{Proposed} & \textbf{0.3912} {} {} \textbf{0.3721}& \textbf{0.2921} {} {} \textbf{0.3024}\\

MRRRec& 0.4220 {} {} 0.3923 & 0.3286 {} {} 0.3762\\
NARRE &  0.5798 {} {} 0.5167 & 0.5165 {} {} 0.4602\\
DAML & 0.6198 {} {} 0.5722& 0.4918 {} {} 0.5013\\
MPCN & 0.5526 {} {} 0.5282 & 0.4526 {} {} 0.5282\\

ANR &0.4431 {} {} 0.4110 &  0.4452 {} {} 0.3921 \\

DeepCoNN & 0.6002 {} {} 0.5535 & 0.5006 {} {} 0.4654\\
Nassar et al.\cite{nassar2020multi} &0.4670 {} {} 0.3945 & 0.3979 {} {} 0.3321\\

\hline

\end{tabular}
\end{center}
\label{tab:baselineError}
\end{table*}

\begin{table*}[t]
\centering
\caption{Performance comparison with baseline in terms of precision, recall, and F1 score}
\begin{center}
\begin{tabular}{p{2.5cm}p{3.8cm}p{3.8cm}}
\hline
\multicolumn{3}{c}{Performance Comparison}\\
\hline
Model & TripAdvisor &  BeerAdvocate\\
\hline
 & Precision{} {} Recall{} {} F1 & Precision{} {} Recall{} {}F1 \\ 
\hline
\textbf{Proposed} & \textbf{0.9511} \textbf{ 0.9596 } \textbf{ 0.9553 } & 0.9095 \textbf{ 0.9421} \textbf{ 0.9255 }\\
MRRRec &0.9301 {} {} 0.9490 {} {} 0.94901 & 0.8938 {} {} 0.9110 {} {} 0.9023 \\
NARRE & 0.8521 {} {} 0.9116 {} {} 0.8808 & 0.8935 {} {} 0.9257{} {} 0.9093 \\
DAML & 0.9010 {} {} 0.8793 {} {} 0.8793 & 0.8968 {} {} 0.8813{} {} 0.8889 \\
MPCN &  0.8323 {} {}  0.9021 {} {} 0.9021 & 0.8776 {} {} 0.9317{} {}  0.9038 \\
ANR & 0.9092 {} {} 0.9521 {} {} 0.9302 & 0.9121 {} {} 0.9001 {} {}   0.9060 \\
DeepCoNN &  0.8725 {} {} 0.9295 {} {} 0.9000 & 0.8582 {} {} 0.9176 {} {} 0.8869 \\
Nassar et al.\cite{nassar2020multi} & 0.8911 {} {} 0.9201	 {} {} 0.9053 & 0.7454 {} {} 0.9247 {} {} 0.8254 \\
\hline

\end{tabular}
\end{center}
\label{tab:baselineAcc}
\end{table*}

\section{Results}
\begin{table*}
\centering
\caption{Rating prediction performance for different scenarios of criteria ratings}
\begin{tabular}{c c c } 
\hline
Criteria & MSE & MAE\\ 
\hline
\textbf{implicit+undefined+predefined}& 0.20532 & 0.32262 \\ 

implicit+predefined&  0.21198 &0.32516 \\ 

undefined+predefined &0.20681& 0.32260\\ 
undefined+implicit &0.78546& 0.69426\\
Only predefined &0.21210& 0.32373\\ 

\hline

\end{tabular}
\label{tab:ablation}
\end{table*}


Table \ref{tab:baselineError} presents the performance comparison of our model with baselines in terms of MSE and MAE. We compare the error measure with seven baselines in a combination of a general recommendation system, a review-based recommendation system, an aspect-based recommendation system, and a multi-criteria recommendation system. We include the baselines that share the source code publicly. For both TripAdvisor and Beer Advocate datasets, our model shows lower MSE and MAE values. The model obtains 3.08\% and 2.02\% lower MSE and MAE respectively compared to the best-performing baseline, i.e., MRRRec on TripAdvisor dataset. For the Beer Advocate dataset, the model achieves a 3.65\% and 2.97\% lower MSE and MAE respectively. The existing state-of-the-art recommendation models use only MSE and MAE for performance evaluation\cite{liu2019daml, nassarmulti, chin2018anr,li2020aspect, shi2019deep} because they use their models for rating prediction tasks. We further extend our evaluations to precision, recall, and F1 scores to understand the recommendation accuracy. In this case, we convert the rating to 0 (not recommend) and 1 (recommend) by considering a threshold rating value of 3.5.

Table \ref{tab:baselineAcc} depicts the accuracy measures of our proposed model and compares them with the baselines. On the Tripadvisor dataset, our model achieves higher accuracy compared to all the baselines in terms of precision, recall, and F1 measures. In comparison to the best-performing baseline, i.e., MRRRec, precision, and recall, and F1 are increased by 0.69\%, 0.35\%, and 0.93\% respectively. For the BeerAdvocate dataset, recall and F1 scores are increased by 1.7\% and 1.95\% respectively. However, precision sees a decrease of 0.25\%, indicating a slightly lower performance compared to the ANR model.

In Table \ref{tab:ablation}, we present the rating prediction performance for different criteria rating scenarios. It is important to note that our focus here is to understand the relationship between different criteria rating sets and the overall rating. It is evident that when we use all three forms of criteria e.g. implicit, undefined, and predefined, the prediction error is the lowest. MSE and MAE are decreased by 0.678\% and 0.11\% respectively. The detailed analysis is provided in Section 5.2.


\subsection{Parameter Sensitivity Analysis}
\begin{figure}[ht]
\centering
\includegraphics[width=0.5\textwidth]{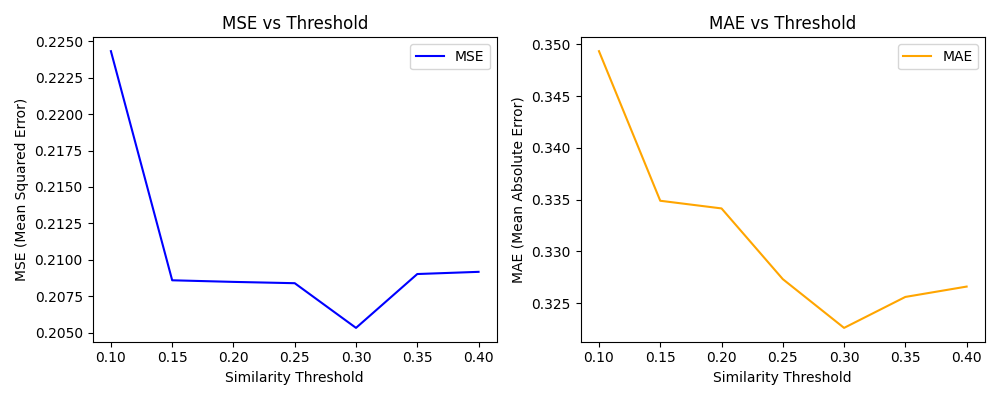}
    \caption{Change of MSE and MAE vs threshold value of similarity score.}
    \label{fig:threshold}
\end{figure}

We evaluate the effectiveness of our model using different threshold values of the similarity score, ranging from 0.10 to 0.40 with an increment of 0.05, as depicted in Figure \ref{fig:threshold}. Both the MSE and MAE are higher for lower threshold values, e.g. 0.10, and optimal MSE and MAE are obtained at 0.30. At a lower threshold score, for each sentence in a given review, the similarity between the sentence and criteria is high, and the majority of the sentences fall into the cluster above that similarity. When similarity is at a higher threshold score, the majority of the sentences fall into the cluster lower than that similarity. Therefore, the model cannot capture all the clusters corresponding to the criteria, and at an optimal threshold score, we obtain optimal clusters.

\subsection{Ablation Study}
We check the necessity of including all types of criteria in overall rating prediction. We analyze the rating prediction performance for five different scenarios: 1) implicit criteria, undefined criteria, and predefined criteria, 2) implicit and predefined criteria, 3) undefined and predefined criteria, 4) undefined criteria and implicit criteria, 5) predefined criteria. The focus in this set of experiments is to study the impact of different criteria ratings on overall ratings. Once we identify the scenario that gives the best performance on predicting overall ratings, we will use that combination of criteria in our final model. The result is shown in Table \ref{tab:ablation}. It is important to note that we conduct the ablation study on TripAdvisor datasets only, because TripAdvisor datasets are the most widely used datasets in multi-criteria rating research\cite{nassarmulti, nassar2020novel, hong2021multi}. For each scenario, we predict criteria ratings and learn the aggregation function for overall rating prediction. Among the five cases, we obtain the best performance when we use all the criteria ratings, e.g. implicit, undefined, and predefined criteria ratings. The worst performance is noted from the undefined and implicit criteria, in which we obtain the MSE value of 0.78546 and the MAE value of 0.69426, respectively. This suggests that relying solely on undefined and implicit criteria extracted from the review is insufficient for capturing the user's overall preferences. Likewise, relying solely on predefined criteria is inadequate, as indicated by MSE and MAE values of 0.21210 and 0.32373, respectively. Therefore, there is a need to combine predefined, undefined, and implicit criteria to learn user preferences.

\subsection{Discussion}
It is challenging to cluster free-form reviews because of the length and structure of the content. The reason we chose the sentence clustering method using the semantic search technique is that other clustering algorithms (e.g. k-means) require setting a predefined number of clusters which is infeasible due to the variable review lengths. We tried to find a similarity score between sentences and criteria at which model gives the best clusters. As we discussed in Section 5.1, we obtained the best performance at a threshold value of 0.30 for the TripAdvisor dataset. Each review contains multiple sentences and it is computationally expensive and time-consuming. Therefore, we limit our experiments to a subset of the TripAdvisor dataset for parameter sensitivity analysis.  

Existing approaches utilize either multiple criteria or reviews to capture the relationship between features (e.g. aspect learning, and criteria ratings) and the overall ratings. We leverage a sentiment analyzer to estimate the criteria ratings from the review. The ratings on predefined criteria and implicit criteria are different. From our results, it is evident that the criteria ratings extracted from reviews have an impact on user preference learning. The results also reveal that it is important to consider users' sentiments on criteria in addition to predefined criteria. Therefore, the performance is highly dependent on the sentiment analyzer. As a future direction, it would be worth exploring the options of developing our approach that can estimate the ratings for each clustered sentence.

Although our method obtained the best performance in terms of both error and accuracy measures for both TripAdvisor and BeerAdvocate datasets, the precision is slightly lower for the BeerAdvocate dataset. This indicates that the similarity threshold is domain-dependent. For the beer review, the threshold value could be different (currently we use the same value 0.3 as the TripAdvisor dataset) which can be explored in future work. It is worth highlighting that recall is generally higher across most models, excluding MRRRec and DAML. This emphasizes the need to delve deeper into the current state-of-the-art accuracy measures.

\section{Conclusion}
We introduce a novel multicriteria recommendation system that utilizes implicit, undefined, and predefined criteria ratings. Implicit and undefined criteria are extracted from the review, using semantic similarity-based sentence clustering techniques. A sentiment analyzer is used to estimate the ratings on these criteria. We show that users' preferences for an item not only depend on explicit ratings given on predefined criteria but also on the implicit and undefined criteria that users express in the review. We employ an aggregation-based method to develop a multi-criteria recommendation system. Predefined, implicit, and undefined criteria are utilized to learn the aggregation function and predict the overall ratings. We show that our proposed method outperforms the existing baselines in terms of both error and accuracy measures on the TripAdvisor dataset and Beer Advocate dataset. In the future, we aim to explore the LLMs to extract the aspects from the review.

\section{Acknowlegements}
This work is partially sponsored by Natural Science and Engineering Research Council of Canada (grant 2020-04760)
The authors would like to thank our reviewers for their tremendous efforts on review and the constructive feedback. 

\balance
\bibliographystyle{IEEEtran}

\bibliography{library}

\vspace{12pt}
\color{red}

\end{document}